\documentclass[10pt]{iopart}

\usepackage{graphicx}

\begin{document}

\title[Domain wall interactions in magnetic tunnel junctions]{Layer-resolved
imaging of domain wall interactions in magnetic tunnel junction-like
trilayers}
\author{J.Vogel$^1$, S.Cherifi$^1$, S.Pizzini$^1$, F.Romanens$^1$, J.Camarero$^2$, F.Petroff$^3$ and
S.Heun$^4$ and A.Locatelli$^5$}
\address{$^1$ Institut N\'{e}el, CNRS \& Universit\'{e} Joseph Fourier, BP166, 38042 Grenoble Cedex 9, France}
\address{$^2$ Dpto. F\'{i}sica de la Materia Condensada, Universidad Aut\'{o}noma de Madrid, 28049
    Madrid, Spain}
\address{$^3$ Unit\'{e} Mixte de Physique CNRS/Thales, Route D\'{e}partementale 128, 91767 Palaiseau Cedex,
France and Universit\'{e} Paris-Sud XI, 91405 Orsay Cedex, France}
\address{$^4$ Laboratorio Nazionale TASC, S.S. 14, km 163.5 in Area Science Park, 34012 Basovizza,
    Trieste, Italy}
\address{$^5$ Sincrotrone Trieste S.C.p.A., S.S. 14, km 163.5 in Area Science Park, 34012 Basovizza,
    Trieste, Italy}


\begin{abstract}We have performed a layer-resolved, microscopic study
of interactions between domain walls in two magnetic layers
separated by a non-magnetic one, using high-resolution x-ray
photoemission electron microscopy. Domain walls in the hard magnetic
Co layer of a Co/Al$_2$O$_3$/FeNi trilayer with in-plane uniaxial
anisotropy strongly modify the local magnetization direction in the
soft magnetic FeNi layer. The stray fields associated to the domain
walls lead to an antiparallel coupling between the local Co and FeNi
moments. For domain walls parallel to the easy magnetization axis
this interaction is limited to the domain wall region itself. For
strongly charged (head-on or tail-to-tail) walls, the antiparallel
coupling dominates the interaction over radial distances up to
several micrometers from the centre of the domain wall.
\end{abstract}

\maketitle

\section{Introduction}
Magnetic trilayer systems consisting of two thin ferromagnetic (FM)
layers separated by a non-magnetic (NM) spacer layer exhibit a
wealth of interesting physical phenomena that are increasingly used
in applications. Examples are the Giant Magneto resistance effect
\cite{Baibich1988,Binasch1989}, the difference in resistance for
parallel and antiparallel alignment of the magnetization directions
of the two FM layers, which is used in magnetic read heads based on
spin-valves \cite{Dieny1991}. The inverse effect, the influence of
an electrical current on the magnetic configuration of the trilayer
system, has drawn a lot of attention recently
\cite{Slonczewski1996,Myers1999,Katine2000}.

In these trilayer systems, interactions between the two FM layers
through the NM spacer layer can be induced either by exchange
\cite{Bruno1992}, magnetostatic \cite{Neel1962} or dynamic effects
\cite{Heinrich2003}. Local variations in these interlayer
interactions, caused by modulated topography \cite{Pennec2004} or
inhomogeneous magnetization, can be important for the static and
dynamic magnetic properties of FM/NM/FM trilayers. In this paper, we
will focus on the effect of domain walls in one or both FM layers,
which lead to strongly localized magnetostatic interactions between
the two layers. These interactions can lead to a local decrease of
tle nucleation barrier for magnetization reversal \cite{Vogel2005b}
or to the demagnetization of one of the layers \cite{Thomas2000a}.
They might also play a role in the decrease of the critical current
in trilayer systems for the propagation of domain walls induced by
spin-polarized current \cite{Grollier2003,Laribi2007}. The goal of
this study was to obtain a better understanding of the interaction
between magnetic layers through a non-magnetic spacer layer in the
vicinity of domain walls, using high-resolution x-ray photoemission
electron microscopy combined with x-ray magnetic circular dichroism
(XMCD-PEEM).

Microscopic evidence of the influence of domain wall stray fields in
one layer on the static domain configuration of another layer was
obtained by different groups
\cite{Fuller1962,Kuch2003,Schafer2002,Schafer2004,Wiebel2005}.
Thomas \textit{et al.} \cite{Thomas2000a} showed that stray fields
of domain walls sweeping back and forth during repeated switching of
one of the FM layers can demagnetise the other magnetic layer, even
if it is much harder magnetically. We have recently shown
\cite{Vogel2005b} that in nanosecond magnetization switching, domain
walls in the hard magnetic Co layer of
Fe$_{20}$Ni$_{80}$/Al$_2$O$_3$/Co trilayers can trigger nucleation
in the soft FeNi layer and thereby increase the local switching
speed. These observations were made taking advantage of the element
selectivity, spatial and temporal resolution of time-resolved
XMCD-PEEM \cite{Kuch2004a,Vogel2003,Bonfim2001,Schneider2006}. The
high-resolution static XMCD-PEEM images of FeNi/Al$_2$O$_3$/Co
trilayers presented in this paper nicely reveal the influence of
stray fields of domain walls in the Co layer on the local
magnetization of the FeNi layer. These stray fields lead to a
localised antiparallel coupling between the Co and FeNi moments. For
strongly charged walls, which are not parallel to the easy
magnetization axis of the films, this antiparallel coupling can
extend over radial distances of several micrometers from the centre
of the domain wall.

\section{Experimental details}

The sample consisted of a stack of continuous layers,
CoO(3nm)/Co(7nm)/Al$_2$O$_3$(2.6nm) /Fe$_{20}$Ni$_{80}$(4nm),
deposited on an oxidized Si(111) substrate. It was covered with 3~nm
of Al to protect the layers from oxidation. The Si(111) substrate
was miscut by $8^{\circ}$ along the [11$\overline{2}$] direction and
then heat-treated to obtain a step-bunched surface, with terraces
having an average width of 20~nm separated by 4~nm high steps
\cite{Sussiau1997}. For sufficiently thin films, this substrate
topography is transmitted to the deposited layers
\cite{Pennec2004,MontaignePhD}. The presence of elongated terraces
leads to a shape-induced uniaxial magnetic anisotropy with the easy
magnetization axis along the long axis of the terraces. The
correlated roughness at the two FM/NM interfaces leads to a
magnetostatic orange-peel coupling \cite{Neel1962} between the two
magnetic layers \cite{Pennec2004,Encinas1999}.

In order to obtain XMCD-PEEM images with high spatial resolution ($<
40 nm$), we made use of the SPELEEM instrument at the
Nanospectroscopy beamline at the synchrotron ELETTRA (Trieste,
Italy) \cite{Locatelli2003,Locatelli2006}. In this microscope, the
x-rays hit the sample surface at a grazing incidence angle of
$16^\circ$. It is therefore particularly suited for the study of
samples with in-plane magnetization. Element-selective magnetic
contrast was obtained using XMCD, with the x-ray energy tuned to the
maximum of the Fe L$_3$ absorption edge (720 eV) for the FeNi layer
and of the Co L$_3$-edge (793 eV) for the Co layer. The images,
obtained at room temperature, represent the asymmetry (difference
divided by sum) of two measurements taken with opposite photon
helicity. This allows optimizing the magnetic contrast while
minimizing topographic effects.

\begin{figure}[b]
\begin{center}
\includegraphics*[bb= 183 381 382 523]{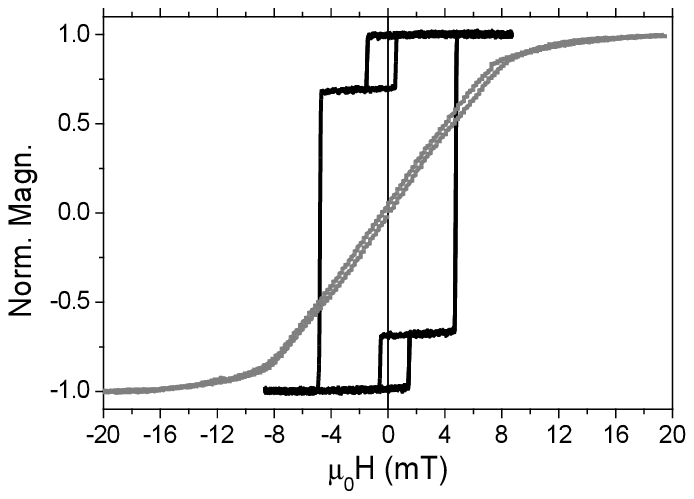}
\caption{Quasi-static hysteresis loops of the magnetic tunnel
junction-like trilayer obtained by longitudinal Kerr effect. Loops
obtained with the field applied along the easy (black line) and hard
(grey line) magnetization axes are shown, as well as minor loops for
the FeNi layer taken along the easy axis.} \label{Hyst}
\end{center}
\end{figure}

\section{Results and discussion}
The quasi-static magnetization hysteresis loops of the sample,
obtained using the longitudinal magneto-optical Kerr effect, are
shown in Fig.~\ref{Hyst}. The quasi-static magnetization reversal of
this sample takes place by nucleation of a small number of domains
and a subsequent propagation of the domain
walls~\cite{Romanens2006}. The minor loops of the FeNi layer reveal
a coupling strength of about 0.6 mT.

\begin{figure}
\begin{center}
\includegraphics*[bb= 190 406 424 545]{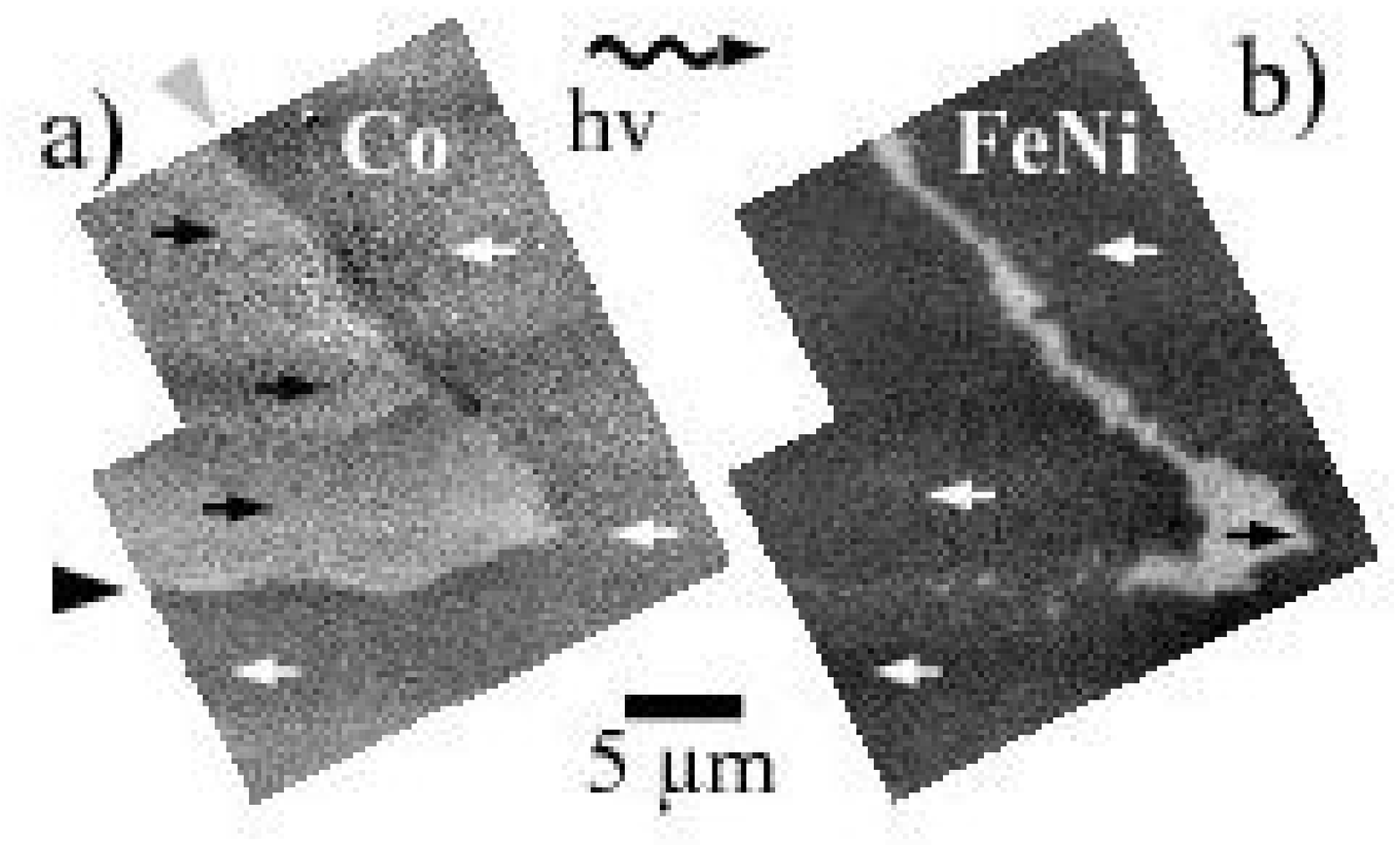}
\caption{Magnetic domain images of the Co (a) and FeNi (b) layers,
after demagnetising the sample and consecutively re-saturating the
FeNi layer. The directions of local magnetization and of incoming
photons are indicated with arrows. The black and grey arrowheads
next to the Co image indicate, respectively, the Co domain wall
oriented parallel to the easy magnetization axis and the one tilted
of about $45^\circ$ with respect to this axis} \label{Easy}
\end{center}
\end{figure}

Before taking images, an AC magnetic field with decreasing amplitude
was applied to the sample along the easy magnetization axis, in
order to create a multidomain state in both the FeNi and Co layers.
A small constant field of about -3 mT was then applied, higher than
the FeNi coercivity but lower than the Co coercivity, in order to
saturate the FeNi layer without changing the Co domain structure
(Fig.~\ref{Hyst}). The sample was then introduced into the
microscope, and several regions of interest were imaged.
Figure~\ref{Easy} shows some typical magnetic domain images of the
Co (a) and FeNi (b) layers. The images were obtained with the
projection of the x-ray incidence direction on the sample surface
parallel to the easy magnetization axis. The contrast is due to the
difference in absorption for domains with their magnetization
direction parallel and antiparallel to the incoming x-ray direction.

The image of the Co layer shows a domain structure typical for a
magnetic layer with in-plane magnetic anisotropy, with one large
white domain (top left part of the image) and one large black
domain. Contrast variations inside the domains are due to
inhomogeneous intensity of the x-ray beam. Two main domain wall
orientations are present, one parallel to the easy magnetization
axis (indicated with a black arrowhead) and the other tilted with
respect to this axis of about $45^\circ$ (grey arrowhead).

\begin{figure}[b]
\begin{center}
\includegraphics*[bb= 171 450 401 503]{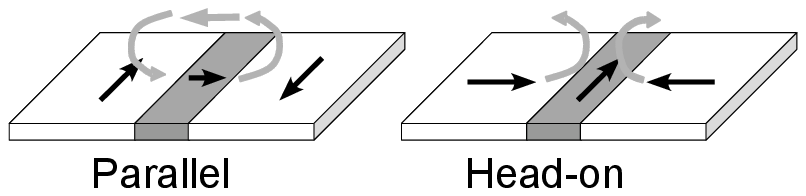} \caption{Schematic
representation of domain wall configurations for `parallel' (left)
and `head-on' (right) N\'{e}el-type domain walls and the associated
domain wall stray fields. The black arrows indicate the local
magnetization direction in the film, the grey arrows the associated
stray field directions.} \label{WallConfig}
\end{center}
\end{figure}

In the FeNi image, clear differences are observed between the
regions above the two sections of the Co domain wall with different
orientations. Above the parallel DW, a faint grey line is visible,
indicating that the magnetization is tilted away from the easy axis
to form a `quasi-wall' \cite{Vogel2005b,Hubertbook}. For the layer
thickness used here, domain walls in both Co and FeNi layers are
expected to be of N\'{e}el-type. The presence of the N\'{e}el wall
in the Co-layer implies a local divergence of the in-plane
magnetization, leading to a magnetic charge $-\nabla \cdot
\textbf{M}$. The associated stray field points in the direction
opposite to the magnetization at the centre of the Co domain wall,
as indicated in Fig.~\ref{WallConfig}(left), and induces the
`quasi-wall' in the FeNi layer.

In the vicinity of the tilted Co DW, much larger white domains are
visible in the FeNi layer. The magnetization directions in the Co
layer around this DW are partly pointing towards each other, leading
to a so-called `head-on' DW. The stray field induced by this
strongly charged DW, as indicated in Fig.~\ref{WallConfig}(right),
is much larger than for the parallel DW. In order to partly
compensate the magnetic charges on the Co DW, an oppositely charged
domain wall is formed in the FeNi layer (a tail-to-tail domain
wall). The Co DW stray field therefore locally induces an {\it
antiparallel} coupling between the two magnetic layers through the
alumina spacer layer, as is clearly seen in Fig.~\ref{Easy}: the
white domains in the FeNi layer are situated on the right side of
the Co DW, above the black domain in the Co layer. This DW induced
antiparallel coupling apparently is much stronger than the
orange-peel coupling that favors parallel alignment.

\begin{figure}
\begin{center}
\includegraphics*[bb= 185 413 418 532]{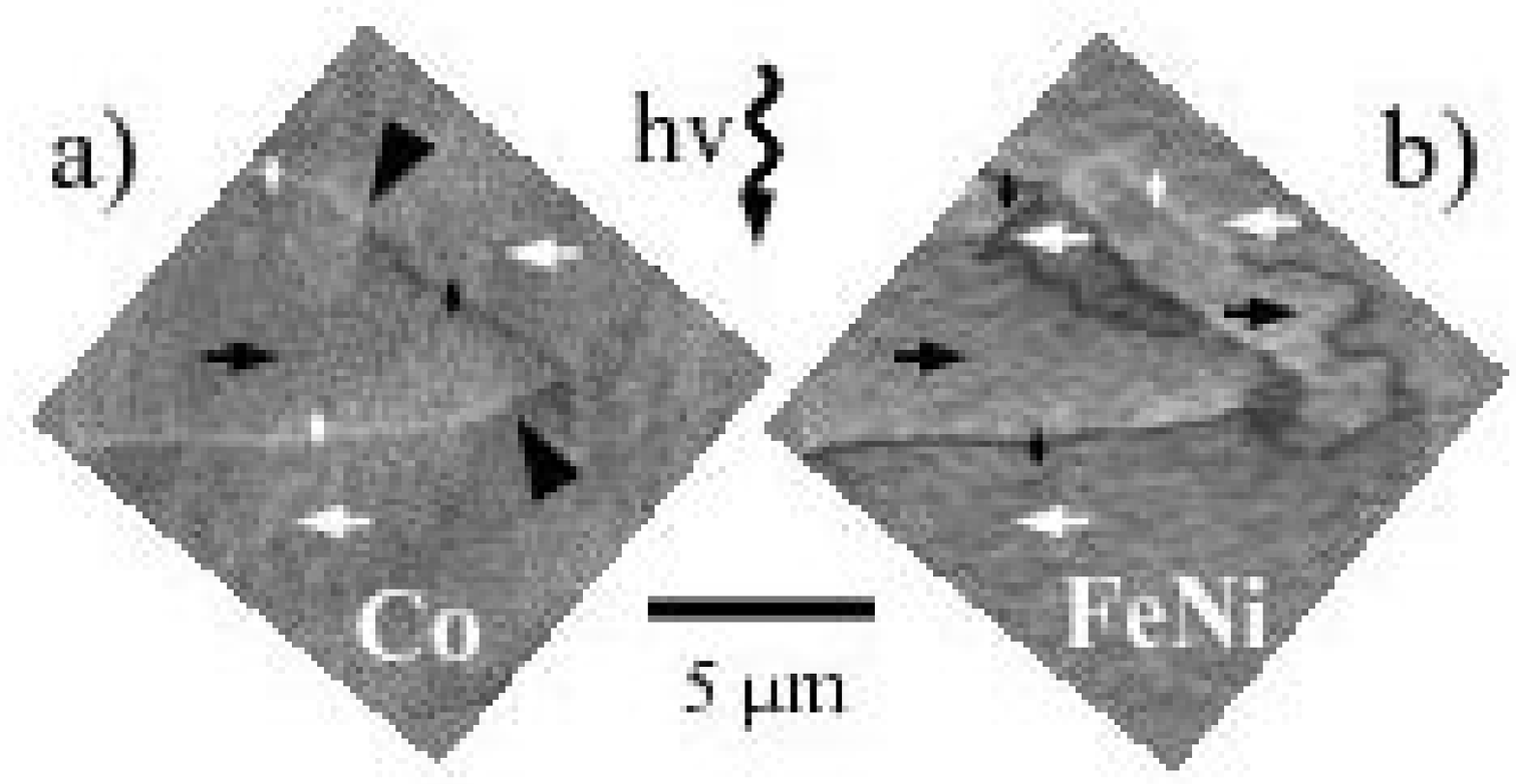}
\caption{Magnetic domain images of the Co (a) and FeNi (b) layers,
taken with the incoming x-rays perpendicular to the easy
magnetization axis. The black and white regions correspond to domain
walls, the grey regions to domains. The black arrowheads in the Co
image indicate the positions of two Bloch lines.} \label{Hard}
\end{center}
\end{figure}

In order to allow a better observation of the domain walls, we
rotated the sample by $90^\circ$. The corresponding images, thus
taken with the x-rays coming in along the hard magnetization axis,
are shown in Fig.~\ref{Hard}. The azimuthal rotation was performed
outside the microscope, since an in-situ rotation was not possible.
This makes it almost impossible to find back exactly the same region
of the sample in this kind of continuous films. The region in
Fig.~\ref{Hard} shows therefore a domain structure that is similar
to the one in Fig.~\ref{Easy}, but not exactly the same. In this
case, the observed black and white regions correspond to the domain
walls or quasi-walls, while the domains show intermediate grey
contrast. In the left (Co) image, a zigzag domain wall is observed
as in Fig.~\ref{Easy}, and the magnetization direction at the centre
of the DW is indicated by the arrows. The expected magnetization
direction inside the domains is also given. Two Bloch lines
(vortices), characterised by a reversal of the direction of
magnetization at the centre of the DW, can be clearly observed. They
are indicated by black arrowheads in Fig.~\ref{Hard}a).

Figure~\ref{Hard}~b) shows that at positions where a domain wall is
present in the Co layer, domain walls or quasi-walls are present
also in the FeNi layer. The contrast at these positions and thus the
magnetization direction at the centre of the wall is opposite in the
Co and FeNi layers. This is due to the direction of the domain wall
stray field, and allows a partial compensation of the domain wall
charges in both layers. Micromagnetic simulations confirm this
picture \cite{Vogel2005b}. Additional domain walls in the FeNi
layer, which are not associated to the Co DW, are due to the
extended domains generated in the FeNi layer by the stray field of
the tilted Co DW (see Fig.~\ref{Easy}b).

Outside the region where domain walls are present, contrast
modulations are visible. These modulations are not due to noise,
since they are reproducible between different series of images taken
at the same position. This indicates that the local magnetization
direction can deviate from the global easy magnetization axis. This
is a remainder of the so-called `ripple' structure often observed
for uniaxial polycrystalline thin films when a magnetic field is
applied perpendicular to the easy magnetization axis. The phenomenon
can be associated to the crystalline anisotropy of individual grains
\cite{Harte1968}. In our case, it is probably associated to a
modulation of the long axis of the substrate terraces with respect
to the easy [110] axis. The long axis of each terrace determines the
local shape anisotropy, leading to a distribution of magnetization
directions around the average, macroscopic easy magnetization axis.

\begin{figure}
\begin{center}
\includegraphics*[bb= 111 397 483 585]{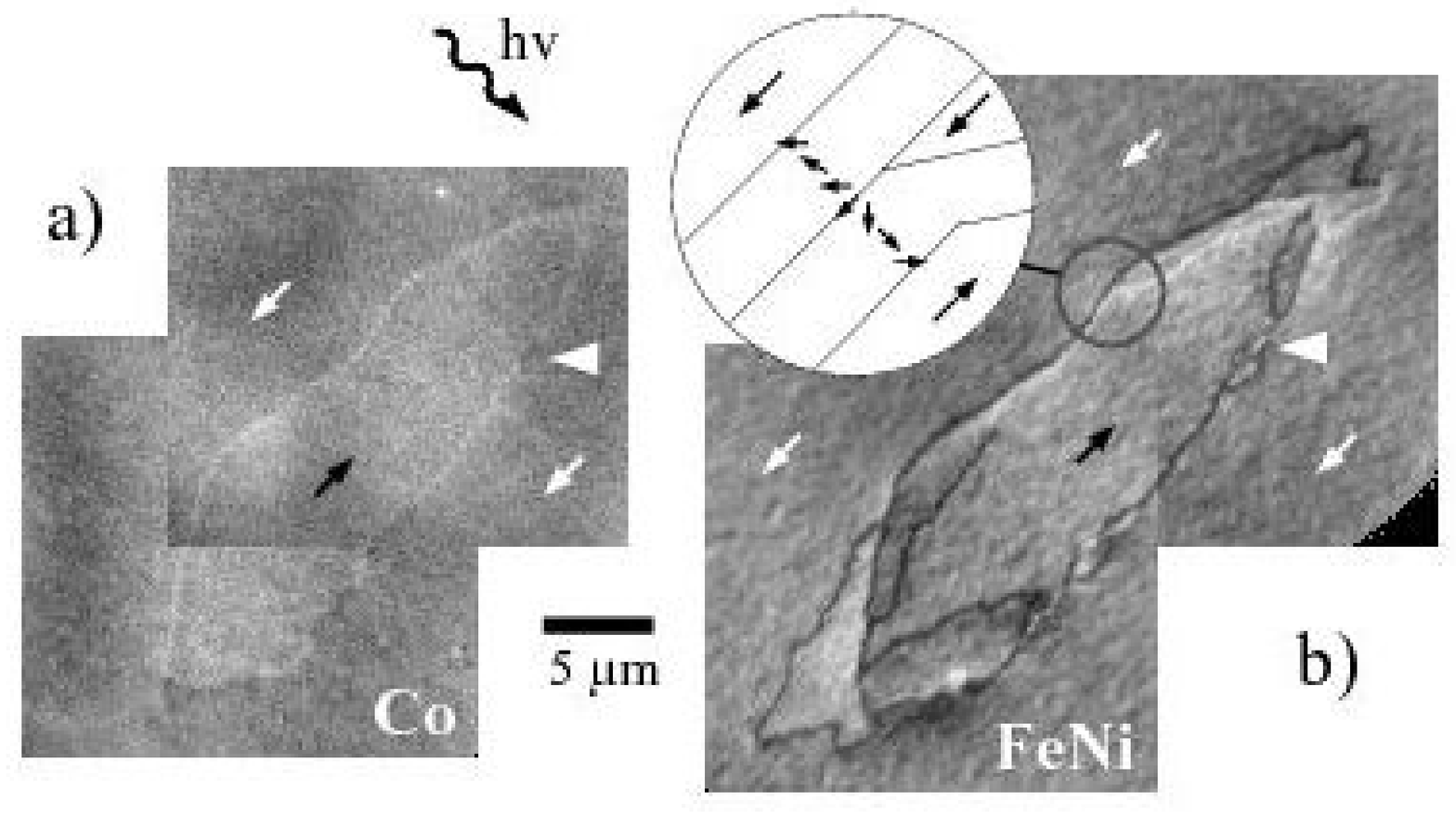} \caption{Magnetic domain
images of the Co (a) and FeNi (b) layers. The images were taken with
the incoming x-ray perpendicular to the easy magnetization axis,
leading to a contrast induced by the domain walls only. The black
circle indicates a position where a $180^\circ$ domain wall and a
quasi-wall join. A schematic representation of the magnetization
directions in this region is shown in the large circle.}
\label{Ellipse}
\end{center}
\end{figure}

In Fig.~\ref{Ellipse}, we show an ellipse-shaped domain structure in
the Co layer, obtained after complete demagnetization of the sample.
In this case, the magnetization directions in both layers are
expected to be globally parallel far from the domain wall regions,
due to the orange-peel coupling. The expected magnetization
directions are indicated by arrows in Fig.~\ref{Ellipse}. In the Co
domain wall, several Bloch lines are clearly visible. At these Bloch
lines, the magnetization direction at the centre of the DW reverses
direction, and therefore also the emitted stray field. In most
cases, the corresponding magnetization direction at the centre of
the FeNi DW also changes sign. This can be seen for instance at the
position indicated by the white arrowhead in Fig.~\ref{Ellipse},
where the small black segment in the Co DW is accompanied by a small
white segment in the FeNi DW. At the centre of the Bloch lines, the
magnetization direction is perpendicular to the plane of the film
\cite{Wachowiak2002}, leading to the presence of additional stray
field components. This leads to a more complex magnetization
configuration in the FeNi film close to the Bloch lines.

The strongly charged domain walls at the extremities of the
ellipse-shaped Co domain induce small domains in the FeNi layer at
both sides of the Co DW. As mentioned above, the magnetic charges in
the Co layer can be partly compensated by creating charges of
opposite sign in the FeNi layer. This leads to a strong
anti-parallel coupling between Co and FeNi layers close to the
charged Co domain walls. This influence of the domain wall charges
is felt up to lateral distances of several micrometers from the
domain wall centre.

At several positions, a grey/black/white/grey contrast can be seen
around domain walls in the FeNi layer. This is what is expected for
a $360^\circ$ domain wall, where the magnetization vector makes a
complete $360^\circ$ turn going from one side of the domain wall to
the other. However, also other configurations can give rise to this
contrast. As an example consider the region in the circle of the
FeNi image of Fig.~\ref{Ellipse}. On the lower left, a black/white
contrast is present. Going to the upper right, the black and white
parts split. The white part is not associated to a DW in the Co
layer, indicating that it has to be a normal $180^\circ$ N\'{e}el
wall. The black part separates two domains with the same
magnetization direction and is therefore a `quasi-wall', induced by
the stray field of the Co domain wall. This means that, crossing the
black/white region from lower right to upper left, the magnetization
vector turns over an angle of about $270^\circ$, before turning back
$-90^\circ$ to end up at an angle of $180^\circ$ with respect to the
initial magnetization direction. The contrast is thus induced by the
combination of a N\'{e}el wall and a `quasi-wall' with the same
chirality. A zoom of the magnetic configuration is schematically
shown in the circle.

These images can be used to obtain an estimate of the domain wall
width in both layers. The definition we use for determining the
domain wall width is based on the total wall flux and is given by
$W_F = \int_{-\infty}^\infty \cos \varphi (x)dx$ \cite{Hubertbook},
where $\varphi(x)$ is the angle between the local magnetization
direction and the hard magnetization axis. This definition is the
most convenient in our case, since the XMCD-intensity in the images
of Fig.~\ref{Hard} is proportional to the projection of the local
magnetization direction on the x-ray incidence direction and thus to
$\cos \varphi$. The wall width can be obtained by taking an image
profile along a direction perpendicular to the DW. Using the highest
resolution images (field of view $10~ \mu m$), we find widths of
$130 \pm 20$ nm for the Co domain walls and $170 \pm 15$ nm for the
FeNi domain walls. However, the domain wall widths seem to be quite
irregular, probably due to the specific, irregular topography of
terraces and steps that can locally lead to confinement or, on the
contrary, largening of the domain walls \cite{Pennec2004}. The DW
width of Co is smaller than for FeNi. According to
Ref.~\cite{Hubertbook}, Chap.~5.5.7, the N\'{e}el wall width in each
layer of a magnetic/non-magnetic/magnetic trilayer system is
determined by the magnetic anisotropy $K_u$ and the stray field
energy (with parameter $K_D=J_S^2/\mu_0$, where $J_S$ is the
spontaneous magnetization). While the magnetic anisotropy tends to
decrease the domain wall width, the stray field energy tends to
increase it. In our case, the magnetic anisotropy is larger for the
Co layer ($K_U^{Co}\approx11200 J/m^3$ against
$K_U^{FeNi}\approx1720 J/m^3$ (Ref.~\cite{Vogel2005b}) but also the
exchange and the stray field energy ($J_S^{Co}=1.76T,
J_S^{FeNi}=1T$). The wall width of the Co domain walls is not much
smaller than the one of the FeNi layer, but the stray field energy
is about three times higher. Moreover, it is easier to deviate the
FeNi magnetization from its easy axis due to the smaller anistropy.
The influence of the Co DW stray field on the FeNi layer is
therefore much larger than the reciprocal effect, which is however
still visible in some images (see Fig.~\ref{Hard}a, for example).

\begin{figure}
\begin{center}
\includegraphics*[bb= 193 421 418 550]{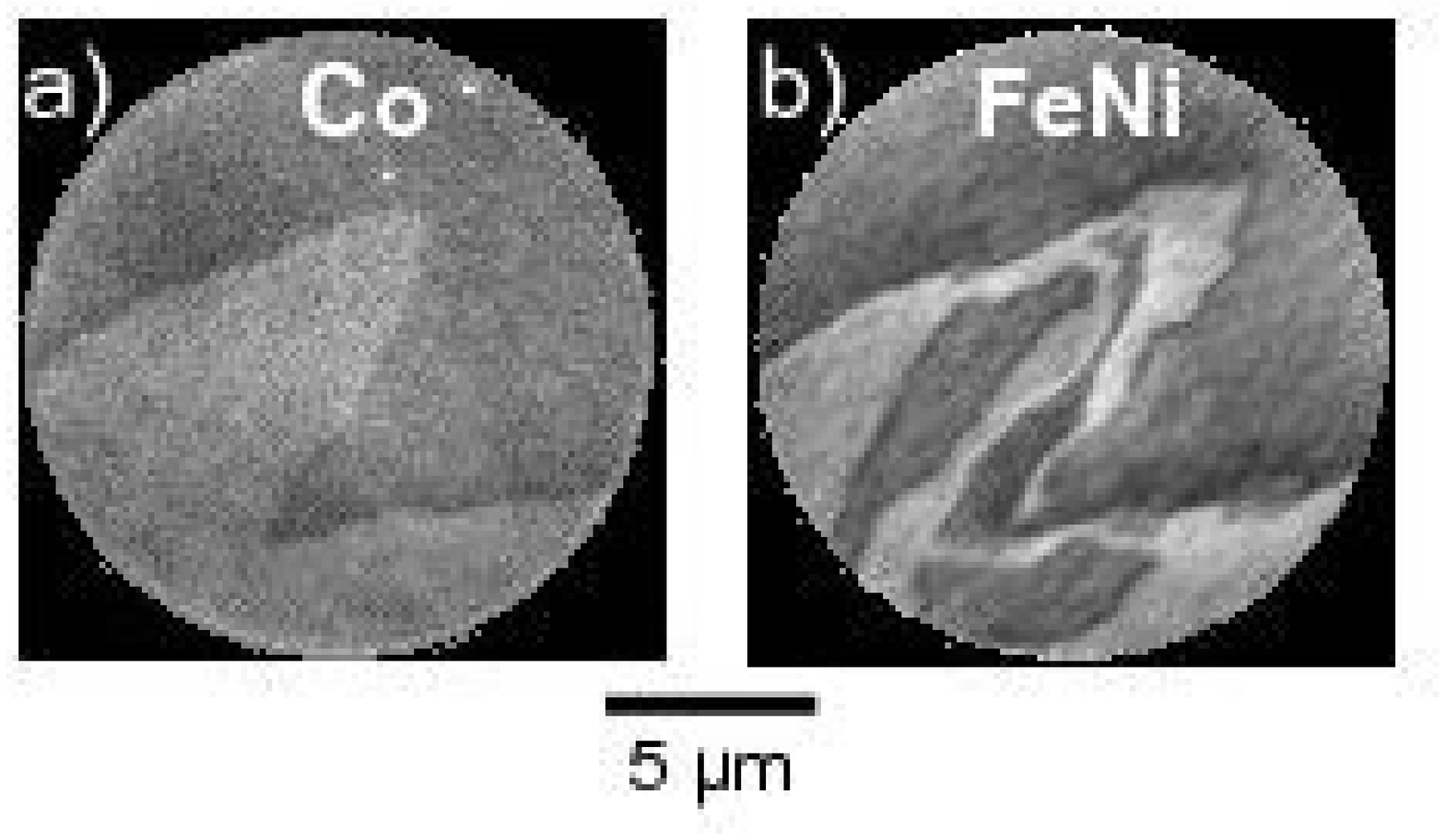}
\caption{Magnetic domain images of the Co (a) and FeNi (b) layers
after application of several bipolar pulses with an amplitude of 5
mT. The images were taken with the x-rays coming in at an azimuthal
angle of $10^\circ$ with respect to the hard magnetization axis.}
\label{Pulses}
\end{center}
\end{figure}

Time-resolved XMCD-PEEM measurements on this same sample already
revealed the strong influence of the Co domain walls on the
magnetization reversal of the FeNi layer (Fig.~7 of
Ref.~\cite{Romanens2006}). In that case, a non-saturated region was
observed in the FeNi layer just above the Co domain wall even for
fields of 6 mT. This region seemed to be rather large (about
1~$\mu$m), larger than we observe in these static images, even
taking into account the difference in spatial resolution (about 50
nm here against 300 nm in Ref.~\cite{Romanens2006}). In order to
investigate the influence of repeated magnetic pulses on the domain
walls, we recorded images of the Co and FeNi layers after
application of some thousands of bipolar magnetic pulses of about
5~mT, using the same coil and pulsed current supply as in
Refs.~\cite{Romanens2006,Vogel2004}. The corresponding images are
shown in Fig.~\ref{Pulses}.

The images were recorded with the x-rays coming in at an azimuthal
angle of $80^\circ$ with respect to the easy magnetization axis. In
this way it is possible to reveal the magnetization direction in
both the domains and the domain walls. The FeNi image clearly shows
signs of domain accumulation around the Co domain wall over
distances well above $1~\mu$m. Compression of these domains during
the application of the field pulses might lead to $360^\circ$ domain
walls and be at the origin of the larger grey region observed in the
images of Ref.~\cite{Romanens2006}.

\section{Conclusion}
We have performed layer-selective imaging of magnetic domains in
FeNi/ Al$_2$O$_3$/Co trilayers with high spatial resolution. These
images reveal the strong interaction between domain walls in the two
magnetic layers. Co domain walls parallel to the easy magnetization
axis induce `quasi-walls' in the FeNi layer, with the magnetization
at the centres of the two walls in opposite directions. For domain
walls that are tilted with respect to the easy magnetization axis
magnetostatic interactions favor an anti-parallel alignment of the
magnetization directions in both magnetic layers over a distance up
to several microns from the domain wall. Upon repetitive application
of magnetic field pulses, these interactions can lead to an
accumulation of domains and domain walls in the FeNi layer around
the Co DW positions.

\section{Acknowledgments} We acknowledge partial funding by the
European Union under contract No. HPRN-CT-2000-00134 and by a
Programme d'Actions Int\'{e}gr\'{e}es `Picasso' (J.V. and J.C.),
through Grant No. HF2003-0173. We thank A. Vaur\`{e}s for her
invaluable help in sample preparation.

\section*{References}
%\begin{thebibliography}{99}

\bibliographystyle{unsrt}
\bibliography{DW_JPCM_final}

%\end{thebibliography}

\end{document}